\documentclass[prx,aps,twocolumn,amsmath,amssymb]{revtex4-1}
\pdfoutput=1
\usepackage{hyperref}
\usepackage{graphicx}

\def\cal#1{\mathcal{#1}}
\def\eqq#1{Eq.~(\ref{#1})}
\def\eq#1{(\ref{#1})}

\def\f#1{Fig.~\ref{#1}}

\def\c#1{~\cite{#1}}

\def\kt{k_{\rm B}T}
\def\e{{\rm e}}

\def\beq{\begin{equation}}
\def\eeq{\end{equation}}
\def\bea{\begin{eqnarray}}
\def\eea{\end{eqnarray}}

\begin{document}
\title{Strong bonds and far-from-equilibrium conditions minimize errors in lattice-gas growth}

\author{Stephen Whitelam}
\email[]{swhitelam@lbl.gov}
\affiliation{Molecular Foundry, Lawrence Berkeley National Laboratory, 1 Cyclotron Road, Berkeley, CA 94720, USA}

\begin{abstract}
We use computer simulation to study the layer-by-layer growth of particle structures in a lattice gas, taking the number of incorporated vacancies as a measure of the quality of the grown structure. By exploiting a dynamic scaling relation between structure quality in and out of equilibrium, we determine that the best quality of structure is obtained, for fixed observation time, with strong interactions and far-from-equilibrium growth conditions. This result contrasts with the usual assumption that weak interactions and mild nonequilibrium conditions are the best way to minimize errors during assembly.
\end{abstract}

\maketitle
{\em Introduction --} Molecular self-assembly is usually done using interaction strengths $\epsilon$ comparable to the thermal energy $\kt$ (henceforth set to unity) and small values of the bulk free-energy difference $\Delta g$ between the structure and the parent phase\c{reinhardt2014numerical,zhang2004self,bianchi2007fully,nykypanchuk2008dna,whitesides2002self,valignat2005reversible}. Small values of $\epsilon$, proportional to the logarithm of the microscopic relaxation time, allow particles to unbind and correct errors during assembly\c{hagan2006dynamic,wilber2007reversible,rapaport2008role,hagan2011mechanisms, whitelam2015statistical}. Small values of $\Delta g$ result in slow growth, allowing more time for this error-correction mechanism to operate. Such mild conditions therefore seem a natural choice for minimizing errors during assembly. Here we show that this expectation is not true of layer-by-layer growth in a three-dimensional (3D) lattice gas, when the vacancy density $\phi$ is used as a measure of the quality of the grown structure. We find that $\phi$ obeys a scaling relationship $(\phi -\phi_{\rm eq})/ \phi_{\rm eq} \propto  \tau_{\rm r}/\tau_{\rm g}$, which contains the equilibrium vacancy density $\phi_{\rm eq}$, and the ratio of the growth timescale $\tau_{\rm g}$ and the microscopic relaxation timescale $\tau_{\rm r}$. For fixed observation time, the highest-quality structures -- i.e. those with fewest vacancies -- are made by using large values of $\epsilon$ and $\Delta g$, and are out of equilibrium. 

This prescription results from a competition between the thermodynamic and dynamic factors present in the scaling relation. Large $\epsilon$ favors few vacancies for two reasons. First, the smallest achievable value of $\phi$ is the equilibrium vacancy density, $\phi_{\rm eq}$, and this decreases exponentially with $\epsilon$ because vacancies are thermally-excited defects. Second, grown structures are {\em more} likely to be in equilibrium (for fixed $\Delta g$) as $\epsilon$ {\em increases}, because the ratio $\tau_{\rm r}/\tau_{\rm g}$ decreases. This is so because layer-by-layer growth proceeds via successive nucleation of 2D layers on a 3D structure\c{burton1951growth,gilmer1976growth,gilmer1980computer,jackson2006kinetic,weeks1979dynamics,de2003principles,sear2007nucleation}. The logarithm of the time for the advance of each layer scales as $\sigma^2/\Delta g$\c{burton1951growth} (where $\sigma \sim \epsilon$\c{onsager1944crystal} is the surface tension between structure and environment), and so grows faster with $\epsilon$ than does the logarithm of the microscopic relaxation time. Thus {\em more} microscopic binding and unbinding events take place during assembly as $\epsilon$ increases at fixed $\Delta g$, and the structure grown is more likely to be in equilibrium. Set against these two factors, as $\epsilon$ increases, larger values of $\Delta g$ are required to produce structures on observable timescales, and as $\Delta g$ increases the ratio $\tau_{\rm r}/\tau_{\rm g}$ increases. For large enough $\Delta g$ we observe the formation of nonequilibrium structures, which contain more vacancies than their equilibrium counterparts. However, this is an acceptable compromise: for fixed observation time the highest-quality structures are obtained for large values of $\epsilon$ and $\Delta g$, such that $\phi_{\rm eq}$ is small and $\tau_{\rm r}/\tau_{\rm g} = {\cal O}(1)$. The structure produced under these conditions is a nonequilibrium one, but is of higher quality than any equilibrium structure that can be grown on comparable timescales. This simple model system therefore defies the expectation that mild nonequilibrium conditions are the best way to minimize errors during assembly.
\begin{figure*}[t]
   \centering
  \includegraphics[width=0.8\linewidth]{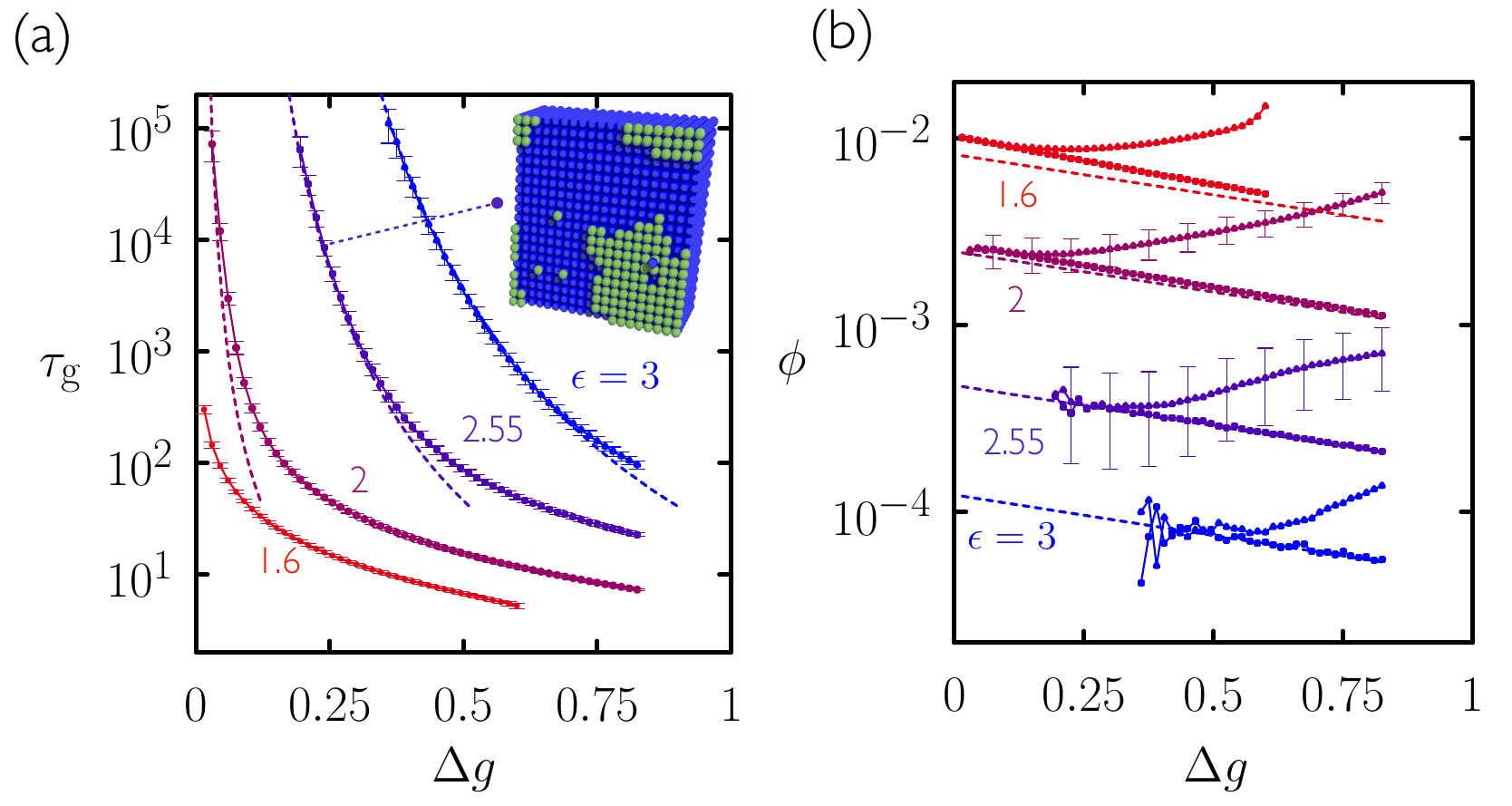} 
   \caption{(a) Characteristic time $\tau_{\rm g}$ to grow a layer of the particle structure, as a function of $\Delta g$, for four values of $\epsilon$. Overlaid dotted lines are the analytic result \eq{eq_tau}. Inset: snapshot of a growing structure, at the indicated state point; green particles are those in the nucleating layer (see \f{fig_s1} for additional detail). (b) Vacancy density $\phi$ as a function of $\Delta g$ for the same four values of $\epsilon$. Upwards-sloping lines with triangle symbols are the values obtained immediately after a structure of 50 layers was grown. Errorbars are shown sparsely, for clarity. Downwards-sloping lines with square symbols are equilibrium results; for small values of $\phi$ these results approach the estimate \eq{est} (dashed lines without symbols).}
   \label{fig1}
\end{figure*}

{\em Model and results --} Consider the 3D Ising lattice gas, which has been used extensively to study crystal growth\c{burton1951growth,gilmer1976growth,weeks1979dynamics,gilmer1980computer,jackson2002interface,jackson2004analytical,jackson2006kinetic,jackson1995monte}. We consider occupied sites to be particles, and unoccupied sites to be vacancies. Nearest-neighbor particles receive an energetic reward of $-\epsilon<0$, and we impose a chemical potential cost $\mu = 3 \epsilon -\Delta g$ for a particle relative to a vacancy (in Ising model language we have coupling $J=\epsilon/4$ and magnetic field $h=\Delta g/2$).  For $\epsilon > 0.886$ we are in the two-phase region, where an interface between the particle phase and the vacancy phase is stable\c{pawley1984monte}. The bulk free-energy difference between particle and vacancy phases, the thermodynamic driving force for growth, is $\Delta g$. Note that the driving force for growth is independent of $\epsilon$: increasing $\epsilon$ makes particles `stickier', but also reduces the effective particle concentration in `solution', $\approx \e^{-3 \epsilon +\Delta g}$.

We used lattices of $L_x \times L_y \times L_z$ sites. For most simulations we set $L_x=L_y=20$ and $L_z=50$. We imposed periodic boundaries in $x$- and $y$ dimensions, and closed boundaries in $z$, which, through choice of initial conditions, is the growth direction. We began each simulation with a layer of particles in the $L_z=0$ plane, in order to study growth without having to wait for nucleation of a 3D structure. We evolved the system using a kinetically constrained grand-canonical Metropolis Monte Carlo algorithm, similar to that used in Refs.\c{whitelam2014self}. At each step we selected at random a lattice site, and proposed a change in state of that site. If the chosen site had fewer than 6 particles as neighbors then we accepted the proposal with probability $\min(1,\e^{-\Delta E})$, where $\Delta E$ is the change in energy resulting from the proposed move. If the chosen site had 6 particles as neighbors then we rejected the move. The purpose of this constraint is to mimic the slow internal relaxation of solid structures: in the absence of the constraint, vacancies internal to the particle structure can simply fill in, which would not happen in a real growth process. This algorithm and model capture in a simple way some of the key physical features of growth, principally that particles can bind and unbind at the growth front, but not within a solid structure. To determine equilibrium we used a standard grand-canonical Metropolis Monte Carlo algorithm with no kinetic constraint. Both constrained and unconstrained algorithms satisfy detailed balance with respect to the same energy function, and so give rise to identical thermodynamics in the long-time limit.
\begin{figure*}[t] 
   \centering
  \includegraphics[width=\linewidth]{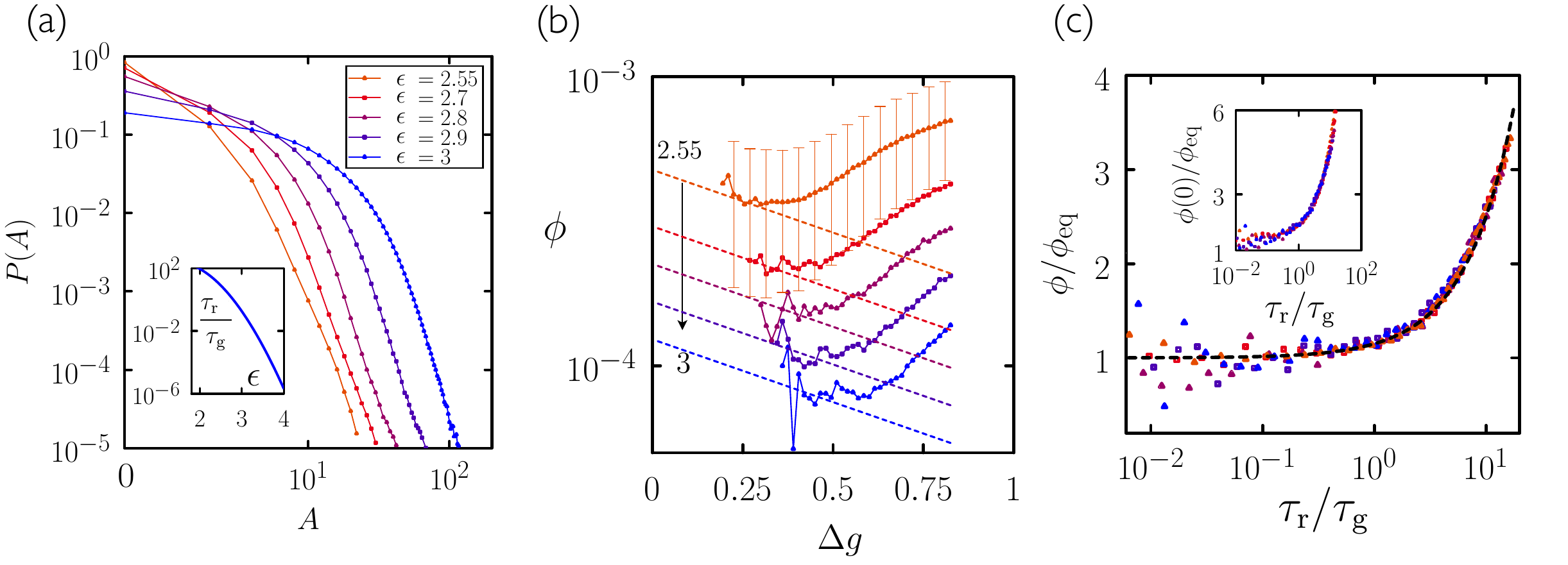} 
   \caption{(a) Probability $P(A)$ that, during growth, a lattice site has undergone $A$ changes of state after first acquiring 6 neighbors. Data are for driving force $\Delta g =0.51$ for various $\epsilon$; the larger is $\epsilon$, the more dynamic are particles on the timescale of growth. Inset: the molecular relaxation time divided by the growth time decreases with increasing $\epsilon$. (b) Dynamic data for large $\epsilon$, in the format of \f{fig1}(b), collapse when rescaled in the manner shown in panel (c). This collapse indicates that grown structures' vacancy densities are controlled by the ratio of relaxation and growth timescales. Inset: the vacancy density $\phi(0)$ of the fresh bulk, scaled by the bulk equilibrium value $\phi_{\rm eq}$, is also a function of $\tau_{\rm r}/\tau_{\rm g}$.}
 \label{fig_scale}
\end{figure*}

For $\Delta g >0$ the particle structure grows in the $z$-direction. In \f{fig1}(a) we show the characteristic time to grow one lattice site in the $z$-direction, averaged over several independent simulations (of order 10 at the smallest values of $\Delta g$, and up to $10^4$ at larger values of $\Delta g$) in which 50 layers were grown. The growth time increases with $\epsilon$ and decreases with $\Delta g$. When $\epsilon$ is small ($\lesssim 2$), the growth front is rough; for larger $\epsilon$ the growth front becomes smooth\c{burton1951growth}. This is the layer-by-layer growth regime. Here it is possible to estimate the growth time by approximating the growth front as a 2D Ising model\c{burton1951growth} (see Appendix \ref{app_a}) and calculating the free-energy barrier (and consequent rate) for the nucleation of successive layers. This can be done analytically using the results of Ryu and Cai\c{ryu2010numerical,ryu2010validity}. Those authors showed that the free-energy cost $G(N)$ for the formation of a cluster of size $N$ in the 2D Ising model can be precisely described by the equation
\beq
\label{ryu}
G(N) = -2 h N +b \sqrt{N} + \tau \ln N + G_0,
\eeq
where $b \equiv 2 \sigma \sqrt{\pi}$ and $G_0 \equiv 8J-b$. The first term in \eq{ryu} is the bulk free-energy reward for growing the stable phase. The term in $\sqrt{N}$ is the cost for creating interface between particles and vacancies; $\sigma$ is the surface tension\c{onsager1944crystal,Shneidman:1999}~\footnote{here $\sigma \equiv (\sigma_\parallel + \sigma_{\rm diag})/(2\sqrt{\chi})$, with $\sigma_\parallel \equiv 2J - \ln \coth J$, $\sigma_{\rm diag} \equiv \sqrt{2} \ln \sinh 2 J$, and $\chi \equiv (1-\sinh^{-4} 2J)^{1/8}$}. These are the usual terms written down in classical nucleation theory (CNT)\c{sear2007nucleation}. The term logarithmic in $N$ (with $\tau=5/4$ in $d=2$) can be interpreted to account for cluster-shape fluctuations. This term is not usually part of a CNT formulation, but is needed to ensure precise agreement with free energies obtained from umbrella sampling\c{ryu2010numerical,ryu2010validity,hedges2012patterning}. The term $G_0$ in \eqq{ryu} ensures that $G(1) = 8J-2h$, which is the free-energy cost for creating one particle in a background of vacancies.
\begin{figure}[b] 
   \centering
  \includegraphics[width=\linewidth]{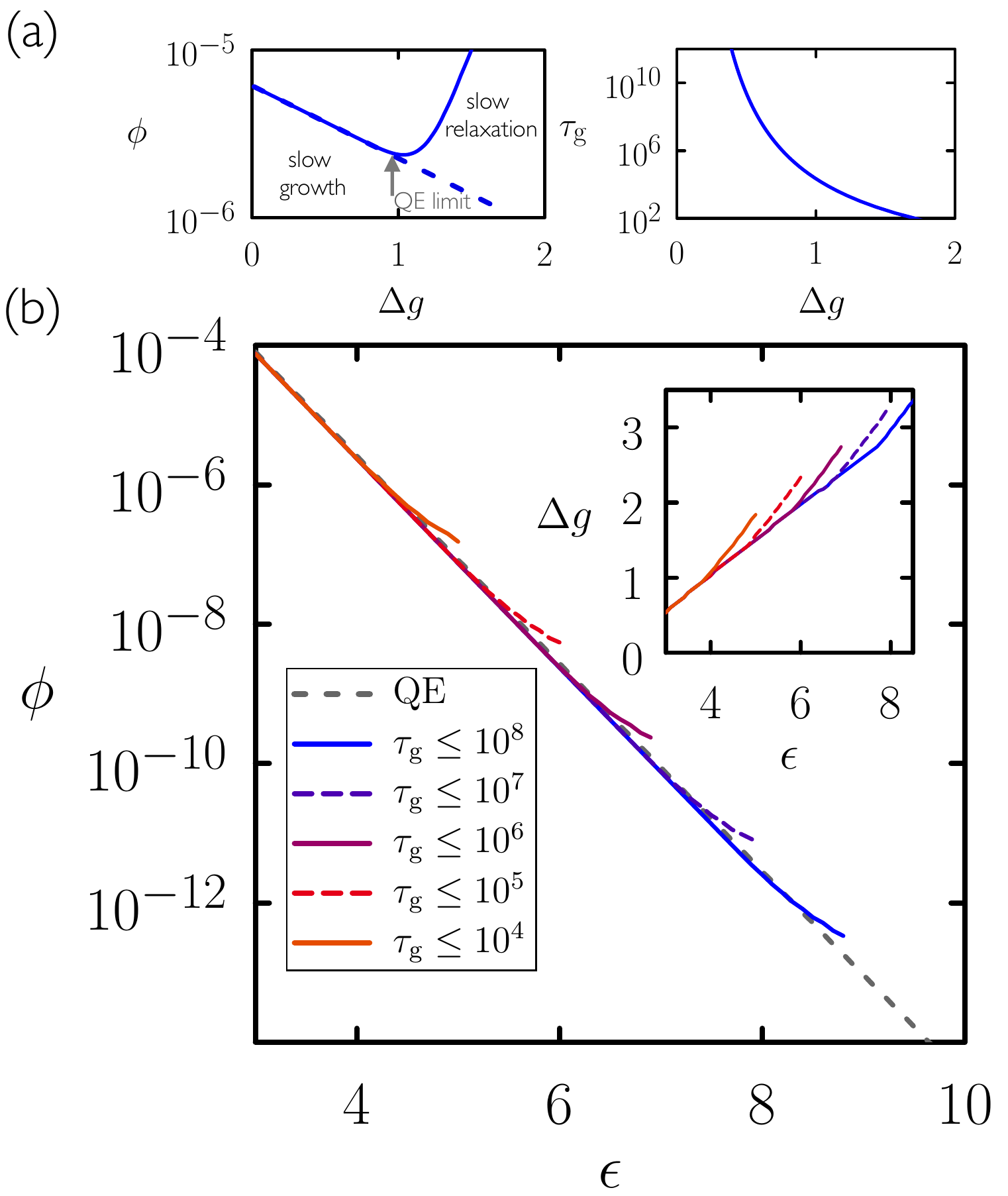} 
   \caption{(a) The scaling relation \eq{scaling} can be used to extrapolate to lengthscales and timescales beyond the reach of simulation (results are for $\epsilon=4$; QE denotes `quasiequilibrium', where the initial outcome of growth is the equilibrium struture). (b) \eqq{scaling} can also be used to determine the protocol that minimizes $\phi$. Each line shows the smallest $\phi$ accessible, as a function of $\epsilon$, for given observation time (we terminate lines when $\tau_{\rm r}/\tau_{\rm g}$ exceeds 20). The inset shows the value of $\Delta g$ required to produce each structure. For each choice of observation time, $\phi$ is minimized for large $\epsilon$ and $\Delta g$, and the structure grown is not an equilibrium one. }
 \label{fig3}
\end{figure}

The critical cluster size $N_{\rm c}$ is the value of $N$ that maximizes \eqq{ryu}, and is $N_{\rm c} = {\cal C} \sigma^2 \pi/(4 h^2)$, where ${\cal C} \equiv (1+\sqrt{1+8 h \tau/\sigma^2 \pi})^2/4$ is a correction, resulting from the logarithmic term, to the usual CNT expression. The free-energy barrier for 2D nucleation is then 
\beq
\label{barrier}
G_{\rm max} = G(N_{\rm c});
\eeq
see \f{fig_barrier}~\footnote{If the cross-sectional area $N_\perp \equiv N_x \times N_y$ of the simulation box is too small to accommodate the 2D critical cluster, $N_\perp < N_{\rm c}$, then \eq{barrier} should be replaced by $G(N_\perp)$.} (in the absence of the logarithmic term, $G_{\rm max}-G_0=\sigma^2 \pi/(2h)$, familiar from CNT). We then estimate the characteristic time for the advance of the growth front to be
\beq
\label{eq_tau}
\tau_{\rm g} = \tau_0 \exp(G_{\rm max}),
\eeq
for sufficiently large $G_{\rm max}$, where $\tau_0=10^{-2}$ is a constant that we determined by comparison with simulation. The estimate \eq{eq_tau} agrees with the simulation data of \f{fig1} for sufficiently large $\epsilon$ and sufficiently small $\Delta g$. This comparison confirms that growth in the regime $\epsilon \gtrsim 2$ is controlled by layer nucleation, and establishes the scaling of growth time with $\epsilon$ for arbitrarily large values of that parameter.

We next assess how close to equilibrium is the structure produced immediately after the growth process. In \f{fig1}(b) we show the vacancy density $\phi$, the number of vacancies divided by the total number of sites, within the middle 50\% of the simulation box (between the planes $z=L_z/4$ and $z=3 L_z/4$) immediately upon completion of layer $L_z=50$. For comparison we show the value of $\phi$ in equilibrium, $\phi_{\rm eq}$. For small values of $\phi_{\rm eq}$ these equilibrium values approach the estimate
\beq
\label{est}
\phi_{\rm eq}^{(0)} = \left(1+\e^{3 \epsilon + \Delta g}\right)^{-1};
\eeq 
note that $6 \epsilon - \mu =3 \epsilon + \Delta g$ is the energy cost for removing a particle from the bulk of a vacancy-free structure.

Comparison of growth and equilibrium results indicates that, for all values of $\epsilon$ studied, there exists for sufficiently small $\Delta g$ a `quasiequilibrium' regime\c{jack2007fluctuation,rapaport2008role,whitelam2015statistical}. Here the initial outcome of growth is the equilibrium structure. The vacancy density of sites that have just acquired 6 neighbors for the first time, which we call the fresh bulk, is not the bulk equilibrium value (see \f{fig_relax}(a)). However, particles in the growth front can unbind, leaving a temporary hole (\f{fig_s1}(a)), and allowing sites in the fresh bulk to change state. For small $\Delta g$, such processes occur enough times that the layers adjacent to the growth front equilibrate before the front moves away.

By contrast, for larger values of $\Delta g$ the bulk of the structure is not in equilibrium. The fresh bulk fails to equilibrate in the presence of the growth front, and becomes trapped out of equilibrium as the front moves away. The timescale for subsequent relaxation to equilibrium is very long, because vacancies, which effectively move by diffusion (see Appendix~\ref{app_vacancies}), cannot catch the ballistically-moving growth front. Nonequilibrium trapping of impurities\c{leamy1980nonequilibrium,kim2008probing} and vacancies\c{van1992nonequilibrium} is seen in crystal growth. Notably, for $\epsilon \gtrsim 2$, the value of $\Delta g$ at which the grown structure falls out of equilibrium {\em increases} with increasing $\epsilon$: `colder' structures are better equilibrated. To understand this result, recall that the growth time scales approximately as the exponential of the free-energy barrier to layer nucleation, or approximately as the exponential of $\epsilon^2$. By contrast, we estimate the microscopic relaxation time $\tau_{\rm r}$ at the growth front (or in the bulk next to a vacancy) to be the characteristic time required to remove a particle with 5 bonds. The energy cost for doing so is $5 \epsilon - \mu =2 \epsilon + \Delta g$, and so we estimate
\beq
\label{rel}
\tau_{\rm r} = \e^{2 \epsilon + \Delta g}.
\eeq
Thus the growth time increases faster with $\epsilon$ than does the relaxation time, and so {\em more} molecular relaxation events take place during growth at large $\epsilon$; see \f{fig_scale}(a).

We can justify the estimate \eq{rel} for relaxation time by rescaling the $\epsilon \gtrsim 2$ data points of \f{fig1}(b) (accompanied, in \f{fig_scale}(b), by additional data) by their equilibrium values, plotted as a function of the ratio of growth time (measured) and relaxation time (\eqq{rel}). We observe the collapse shown in \f{fig_scale}(c). This collapse indicates that the nonequilibrium vacancy density is controlled by the ratio of growth and relaxation times; note that collapsed data involve values of $\phi$ that differ by about an order of magnitude, and growth times $\tau_{\rm g}$ that differ by several orders of magnitude. Such dynamic scaling is also seen in simulations of crystal growth in the presence of impurities\c{jackson2002interface,jackson2004analytical,kim2008probing}, vapor deposition of glasses\c{berthier2017create}, irreversible polymerization\c{corezzi2009connecting}, and the growth of model 1D structures\c{whitelam2012self}. 

The black dotted line in \f{fig_scale}(c) has equation
\beq
\label{scaling}
\phi = \phi_{\rm eq} \left(1+k \frac{\tau_{\rm r}}{\tau_{\rm g}} \right),
\eeq
with $k = 0.15$. This expression emphasizes that the outcome of self-assembly is a combination of thermodynamics and dynamics. It also shows how a quasiequilibrium regime, for which $\phi = \phi_{\rm eq}$, emerges when driving is weak. As $\Delta g$ is made small, the growth time diverges -- it scales to leading order as $\exp(\sigma^2 \pi/\Delta g)$ -- while the molecular relaxation time approaches a constant, rendering $\tau_{\rm r}/\tau_{\rm g} \approx 0$. By contrast, for the growing two-component fiber of Ref.\c{whitelam2012self} there is no quasiequilibrium regime, because growth and relaxation times remain strongly coupled even for weak driving. These distinct behaviors indicate an important difference between growth processes in 1D and 3D.

For large $\epsilon$ the quantities $\phi_{\rm eq}$, $\tau_{\rm g}$ and $\tau_{\rm r}$ are accurately described by Equations \eq{est}, \eq{eq_tau}, and \eq{rel}, respectively, and in that regime we can use \eq{scaling} to extrapolate analytically the data of \f{fig1} to lengthscales $\phi^{-1/3}$ and timescales $\tau_{\rm g}$ beyond those accessible to simulation: see \f{fig3}(a) and \f{fig1_extrap}. We can also use it to determine the protocol for producing the highest-quality structure. In \f{fig3}(b) we show the smallest value of $\phi$, as a function of $\epsilon$, accessible on a given observation time (the inset shows the corresponding value of $\Delta g$). In all cases $\phi$ is minimized by large values of $\epsilon$ and values of $\Delta g$ large enough that the structure grown is a nonequilibrium one. In essence, the prescription for the best-quality structure is to have $\epsilon$ large, so that $\phi_{\rm eq}$ is small, and drive the system hard so that the structure grows on the accessible timescale. Consequently, $\tau_{\rm r}/\tau_{\rm g} \gtrsim 1$, meaning that growth is far from equilibrium and results in a nonequilibrium structure.

{\em Conclusions --} The majority of self-assembled materials made with few defects are prepared using weak interactions and mild nonequilibrium conditions, but we have shown that vacancy incorporation in the layer-by-layer growth of a 3D lattice gas is minimized using strong interactions and far-from-equilibrium conditions. Finding error-minimization protocols is important for the assembly of certain types of nanomaterials. For instance, DNA bricks are distinguishable structures built from $Q$ `brick' types, in which each brick possesses a defined location\c{ke2012three,reinhardt2014numerical}. The interaction energies of bricks must grow as $\epsilon \sim \ln Q$ in order to thermally stabilize the assembly (to counter the entropy of permutation $\ln Q!$ possessed by disordered arrangements of bricks). The present work suggests one way to incorporate strong interactions into a productive assembly protocol.

{\em Acknowledgments --} I thank Jeremy D. Schmit for valuable discussions and comments on the manuscript. This work was done at the Molecular Foundry, Lawrence Berkeley National Laboratory, and was supported by the Office of Science, Office of Basic Energy Sciences, of the U.S. Department of Energy under Contract No. DE-AC02--05CH11231.


\begin{thebibliography}{40}%
\makeatletter
\providecommand \@ifxundefined [1]{%
 \@ifx{#1\undefined}
}%
\providecommand \@ifnum [1]{%
 \ifnum #1\expandafter \@firstoftwo
 \else \expandafter \@secondoftwo
 \fi
}%
\providecommand \@ifx [1]{%
 \ifx #1\expandafter \@firstoftwo
 \else \expandafter \@secondoftwo
 \fi
}%
\providecommand \natexlab [1]{#1}%
\providecommand \enquote  [1]{``#1''}%
\providecommand \bibnamefont  [1]{#1}%
\providecommand \bibfnamefont [1]{#1}%
\providecommand \citenamefont [1]{#1}%
\providecommand \href@noop [0]{\@secondoftwo}%
\providecommand \href [0]{\begingroup \@sanitize@url \@href}%
\providecommand \@href[1]{\@@startlink{#1}\@@href}%
\providecommand \@@href[1]{\endgroup#1\@@endlink}%
\providecommand \@sanitize@url [0]{\catcode `\\12\catcode `\$12\catcode
  `\&12\catcode `\#12\catcode `\^12\catcode `\_12\catcode `\%12\relax}%
\providecommand \@@startlink[1]{}%
\providecommand \@@endlink[0]{}%
\providecommand \url  [0]{\begingroup\@sanitize@url \@url }%
\providecommand \@url [1]{\endgroup\@href {#1}{\urlprefix }}%
\providecommand \urlprefix  [0]{URL }%
\providecommand \Eprint [0]{\href }%
\providecommand \doibase [0]{http://dx.doi.org/}%
\providecommand \selectlanguage [0]{\@gobble}%
\providecommand \bibinfo  [0]{\@secondoftwo}%
\providecommand \bibfield  [0]{\@secondoftwo}%
\providecommand \translation [1]{[#1]}%
\providecommand \BibitemOpen [0]{}%
\providecommand \bibitemStop [0]{}%
\providecommand \bibitemNoStop [0]{.\EOS\space}%
\providecommand \EOS [0]{\spacefactor3000\relax}%
\providecommand \BibitemShut  [1]{\csname bibitem#1\endcsname}%
\let\auto@bib@innerbib\@empty
\bibitem [{\citenamefont {Reinhardt}\ and\ \citenamefont
  {Frenkel}(2014)}]{reinhardt2014numerical}%
  \BibitemOpen
  \bibfield  {author} {\bibinfo {author} {\bibfnamefont {A.}~\bibnamefont
  {Reinhardt}}\ and\ \bibinfo {author} {\bibfnamefont {D.}~\bibnamefont
  {Frenkel}},\ }\href@noop {} {\bibfield  {journal} {\bibinfo  {journal}
  {Physical Review Letters}\ }\textbf {\bibinfo {volume} {112}},\ \bibinfo
  {pages} {238103} (\bibinfo {year} {2014})}\BibitemShut {NoStop}%
\bibitem [{\citenamefont {Zhang}\ and\ \citenamefont
  {Glotzer}(2004)}]{zhang2004self}%
  \BibitemOpen
  \bibfield  {author} {\bibinfo {author} {\bibfnamefont {Z.}~\bibnamefont
  {Zhang}}\ and\ \bibinfo {author} {\bibfnamefont {S.~C.}\ \bibnamefont
  {Glotzer}},\ }\href@noop {} {\bibfield  {journal} {\bibinfo  {journal} {Nano
  Letters}\ }\textbf {\bibinfo {volume} {4}},\ \bibinfo {pages} {1407}
  (\bibinfo {year} {2004})}\BibitemShut {NoStop}%
\bibitem [{\citenamefont {Bianchi}\ \emph {et~al.}(2007)\citenamefont
  {Bianchi}, \citenamefont {Tartaglia}, \citenamefont {La~Nave},\ and\
  \citenamefont {Sciortino}}]{bianchi2007fully}%
  \BibitemOpen
  \bibfield  {author} {\bibinfo {author} {\bibfnamefont {E.}~\bibnamefont
  {Bianchi}}, \bibinfo {author} {\bibfnamefont {P.}~\bibnamefont {Tartaglia}},
  \bibinfo {author} {\bibfnamefont {E.}~\bibnamefont {La~Nave}}, \ and\
  \bibinfo {author} {\bibfnamefont {F.}~\bibnamefont {Sciortino}},\ }\href@noop
  {} {\bibfield  {journal} {\bibinfo  {journal} {The Journal of Physical
  Chemistry B}\ }\textbf {\bibinfo {volume} {111}},\ \bibinfo {pages} {11765}
  (\bibinfo {year} {2007})}\BibitemShut {NoStop}%
\bibitem [{\citenamefont {Nykypanchuk}\ \emph {et~al.}(2008)\citenamefont
  {Nykypanchuk}, \citenamefont {Maye}, \citenamefont {Van Der~Lelie},\ and\
  \citenamefont {Gang}}]{nykypanchuk2008dna}%
  \BibitemOpen
  \bibfield  {author} {\bibinfo {author} {\bibfnamefont {D.}~\bibnamefont
  {Nykypanchuk}}, \bibinfo {author} {\bibfnamefont {M.~M.}\ \bibnamefont
  {Maye}}, \bibinfo {author} {\bibfnamefont {D.}~\bibnamefont {Van Der~Lelie}},
  \ and\ \bibinfo {author} {\bibfnamefont {O.}~\bibnamefont {Gang}},\
  }\href@noop {} {\bibfield  {journal} {\bibinfo  {journal} {Nature}\ }\textbf
  {\bibinfo {volume} {451}},\ \bibinfo {pages} {549} (\bibinfo {year}
  {2008})}\BibitemShut {NoStop}%
\bibitem [{\citenamefont {Whitesides}\ and\ \citenamefont
  {Grzybowski}(2002)}]{whitesides2002self}%
  \BibitemOpen
  \bibfield  {author} {\bibinfo {author} {\bibfnamefont {G.~M.}\ \bibnamefont
  {Whitesides}}\ and\ \bibinfo {author} {\bibfnamefont {B.}~\bibnamefont
  {Grzybowski}},\ }\href@noop {} {\bibfield  {journal} {\bibinfo  {journal}
  {Science}\ }\textbf {\bibinfo {volume} {295}},\ \bibinfo {pages} {2418}
  (\bibinfo {year} {2002})}\BibitemShut {NoStop}%
\bibitem [{\citenamefont {Valignat}\ \emph {et~al.}(2005)\citenamefont
  {Valignat}, \citenamefont {Theodoly}, \citenamefont {Crocker}, \citenamefont
  {Russel},\ and\ \citenamefont {Chaikin}}]{valignat2005reversible}%
  \BibitemOpen
  \bibfield  {author} {\bibinfo {author} {\bibfnamefont {M.-P.}\ \bibnamefont
  {Valignat}}, \bibinfo {author} {\bibfnamefont {O.}~\bibnamefont {Theodoly}},
  \bibinfo {author} {\bibfnamefont {J.~C.}\ \bibnamefont {Crocker}}, \bibinfo
  {author} {\bibfnamefont {W.~B.}\ \bibnamefont {Russel}}, \ and\ \bibinfo
  {author} {\bibfnamefont {P.~M.}\ \bibnamefont {Chaikin}},\ }\href@noop {}
  {\bibfield  {journal} {\bibinfo  {journal} {Proceedings of the National
  Academy of Sciences of the United States of America}\ }\textbf {\bibinfo
  {volume} {102}},\ \bibinfo {pages} {4225} (\bibinfo {year}
  {2005})}\BibitemShut {NoStop}%
\bibitem [{\citenamefont {Hagan}\ and\ \citenamefont
  {Chandler}(2006)}]{hagan2006dynamic}%
  \BibitemOpen
  \bibfield  {author} {\bibinfo {author} {\bibfnamefont {M.~F.}\ \bibnamefont
  {Hagan}}\ and\ \bibinfo {author} {\bibfnamefont {D.}~\bibnamefont
  {Chandler}},\ }\href@noop {} {\bibfield  {journal} {\bibinfo  {journal}
  {Biophysical Journal}\ }\textbf {\bibinfo {volume} {91}},\ \bibinfo {pages}
  {42} (\bibinfo {year} {2006})}\BibitemShut {NoStop}%
\bibitem [{\citenamefont {Wilber}\ \emph {et~al.}(2007)\citenamefont {Wilber},
  \citenamefont {Doye}, \citenamefont {Louis}, \citenamefont {Noya},
  \citenamefont {Miller},\ and\ \citenamefont {Wong}}]{wilber2007reversible}%
  \BibitemOpen
  \bibfield  {author} {\bibinfo {author} {\bibfnamefont {A.~W.}\ \bibnamefont
  {Wilber}}, \bibinfo {author} {\bibfnamefont {J.~P.}\ \bibnamefont {Doye}},
  \bibinfo {author} {\bibfnamefont {A.~A.}\ \bibnamefont {Louis}}, \bibinfo
  {author} {\bibfnamefont {E.~G.}\ \bibnamefont {Noya}}, \bibinfo {author}
  {\bibfnamefont {M.~A.}\ \bibnamefont {Miller}}, \ and\ \bibinfo {author}
  {\bibfnamefont {P.}~\bibnamefont {Wong}},\ }\href@noop {} {\bibfield
  {journal} {\bibinfo  {journal} {The Journal of Chemical Physics}\ }\textbf
  {\bibinfo {volume} {127}},\ \bibinfo {pages} {085106} (\bibinfo {year}
  {2007})}\BibitemShut {NoStop}%
\bibitem [{\citenamefont {Rapaport}(2008)}]{rapaport2008role}%
  \BibitemOpen
  \bibfield  {author} {\bibinfo {author} {\bibfnamefont {D.}~\bibnamefont
  {Rapaport}},\ }\href@noop {} {\bibfield  {journal} {\bibinfo  {journal}
  {Physical Review Letters}\ }\textbf {\bibinfo {volume} {101}},\ \bibinfo
  {pages} {186101} (\bibinfo {year} {2008})}\BibitemShut {NoStop}%
\bibitem [{\citenamefont {Hagan}\ \emph {et~al.}(2011)\citenamefont {Hagan},
  \citenamefont {Elrad},\ and\ \citenamefont {Jack}}]{hagan2011mechanisms}%
  \BibitemOpen
  \bibfield  {author} {\bibinfo {author} {\bibfnamefont {M.}~\bibnamefont
  {Hagan}}, \bibinfo {author} {\bibfnamefont {O.}~\bibnamefont {Elrad}}, \ and\
  \bibinfo {author} {\bibfnamefont {R.}~\bibnamefont {Jack}},\ }\href@noop {}
  {\bibfield  {journal} {\bibinfo  {journal} {The Journal of Chemical Physics}\
  }\textbf {\bibinfo {volume} {135}},\ \bibinfo {pages} {104115} (\bibinfo
  {year} {2011})}\BibitemShut {NoStop}%
\bibitem [{\citenamefont {Whitelam}\ and\ \citenamefont
  {Jack}(2015)}]{whitelam2015statistical}%
  \BibitemOpen
  \bibfield  {author} {\bibinfo {author} {\bibfnamefont {S.}~\bibnamefont
  {Whitelam}}\ and\ \bibinfo {author} {\bibfnamefont {R.~L.}\ \bibnamefont
  {Jack}},\ }\href@noop {} {\bibfield  {journal} {\bibinfo  {journal} {Annual
  Review of physical chemistry}\ }\textbf {\bibinfo {volume} {66}},\ \bibinfo
  {pages} {143} (\bibinfo {year} {2015})}\BibitemShut {NoStop}%
\bibitem [{\citenamefont {Burton}\ \emph {et~al.}(1951)\citenamefont {Burton},
  \citenamefont {Cabrera},\ and\ \citenamefont {Frank}}]{burton1951growth}%
  \BibitemOpen
  \bibfield  {author} {\bibinfo {author} {\bibfnamefont {W.-K.}\ \bibnamefont
  {Burton}}, \bibinfo {author} {\bibfnamefont {N.}~\bibnamefont {Cabrera}}, \
  and\ \bibinfo {author} {\bibfnamefont {F.}~\bibnamefont {Frank}},\
  }\href@noop {} {\bibfield  {journal} {\bibinfo  {journal} {Philosophical
  Transactions of the Royal Society of London A: Mathematical, Physical and
  Engineering Sciences}\ }\textbf {\bibinfo {volume} {243}},\ \bibinfo {pages}
  {299} (\bibinfo {year} {1951})}\BibitemShut {NoStop}%
\bibitem [{\citenamefont {Gilmer}(1976)}]{gilmer1976growth}%
  \BibitemOpen
  \bibfield  {author} {\bibinfo {author} {\bibfnamefont {G.}~\bibnamefont
  {Gilmer}},\ }\href@noop {} {\bibfield  {journal} {\bibinfo  {journal}
  {Journal of Crystal Growth}\ }\textbf {\bibinfo {volume} {36}},\ \bibinfo
  {pages} {15} (\bibinfo {year} {1976})}\BibitemShut {NoStop}%
\bibitem [{\citenamefont {Gilmer}(1980)}]{gilmer1980computer}%
  \BibitemOpen
  \bibfield  {author} {\bibinfo {author} {\bibfnamefont {G.}~\bibnamefont
  {Gilmer}},\ }\href@noop {} {\bibfield  {journal} {\bibinfo  {journal}
  {Science}\ }\textbf {\bibinfo {volume} {208}},\ \bibinfo {pages} {355}
  (\bibinfo {year} {1980})}\BibitemShut {NoStop}%
\bibitem [{\citenamefont {Jackson}(2006)}]{jackson2006kinetic}%
  \BibitemOpen
  \bibfield  {author} {\bibinfo {author} {\bibfnamefont {K.~A.}\ \bibnamefont
  {Jackson}},\ }\href@noop {} {\emph {\bibinfo {title} {Kinetic Processes:
  Crystal Growth, Diffusion, and Phase Transformations in Materials}}}\
  (\bibinfo  {publisher} {John Wiley \& Sons},\ \bibinfo {year}
  {2006})\BibitemShut {NoStop}%
\bibitem [{\citenamefont {Weeks}\ and\ \citenamefont
  {Gilmer}(1979)}]{weeks1979dynamics}%
  \BibitemOpen
  \bibfield  {author} {\bibinfo {author} {\bibfnamefont {J.~D.}\ \bibnamefont
  {Weeks}}\ and\ \bibinfo {author} {\bibfnamefont {G.~H.}\ \bibnamefont
  {Gilmer}},\ }\href@noop {} {\bibfield  {journal} {\bibinfo  {journal} {Adv.
  Chem. Phys}\ }\textbf {\bibinfo {volume} {40}},\ \bibinfo {pages} {157}
  (\bibinfo {year} {1979})}\BibitemShut {NoStop}%
\bibitem [{\citenamefont {De~Yoreo}\ and\ \citenamefont
  {Vekilov}(2003)}]{de2003principles}%
  \BibitemOpen
  \bibfield  {author} {\bibinfo {author} {\bibfnamefont {J.~J.}\ \bibnamefont
  {De~Yoreo}}\ and\ \bibinfo {author} {\bibfnamefont {P.~G.}\ \bibnamefont
  {Vekilov}},\ }\href@noop {} {\bibfield  {journal} {\bibinfo  {journal}
  {Reviews in mineralogy and geochemistry}\ }\textbf {\bibinfo {volume} {54}},\
  \bibinfo {pages} {57} (\bibinfo {year} {2003})}\BibitemShut {NoStop}%
\bibitem [{\citenamefont {Sear}(2007)}]{sear2007nucleation}%
  \BibitemOpen
  \bibfield  {author} {\bibinfo {author} {\bibfnamefont {R.~P.}\ \bibnamefont
  {Sear}},\ }\href@noop {} {\bibfield  {journal} {\bibinfo  {journal} {Journal
  of Physics: Condensed Matter}\ }\textbf {\bibinfo {volume} {19}},\ \bibinfo
  {pages} {033101} (\bibinfo {year} {2007})}\BibitemShut {NoStop}%
\bibitem [{\citenamefont {Onsager}(1944)}]{onsager1944crystal}%
  \BibitemOpen
  \bibfield  {author} {\bibinfo {author} {\bibfnamefont {L.}~\bibnamefont
  {Onsager}},\ }\href@noop {} {\bibfield  {journal} {\bibinfo  {journal}
  {Physical Review}\ }\textbf {\bibinfo {volume} {65}},\ \bibinfo {pages} {117}
  (\bibinfo {year} {1944})}\BibitemShut {NoStop}%
\bibitem [{\citenamefont {Jackson}(2002)}]{jackson2002interface}%
  \BibitemOpen
  \bibfield  {author} {\bibinfo {author} {\bibfnamefont {K.~A.}\ \bibnamefont
  {Jackson}},\ }\href@noop {} {\bibfield  {journal} {\bibinfo  {journal}
  {Interface Science}\ }\textbf {\bibinfo {volume} {10}},\ \bibinfo {pages}
  {159} (\bibinfo {year} {2002})}\BibitemShut {NoStop}%
\bibitem [{\citenamefont {Jackson}\ \emph {et~al.}(2004)\citenamefont
  {Jackson}, \citenamefont {Beatty},\ and\ \citenamefont
  {Gudgel}}]{jackson2004analytical}%
  \BibitemOpen
  \bibfield  {author} {\bibinfo {author} {\bibfnamefont {K.~A.}\ \bibnamefont
  {Jackson}}, \bibinfo {author} {\bibfnamefont {K.~M.}\ \bibnamefont {Beatty}},
  \ and\ \bibinfo {author} {\bibfnamefont {K.~A.}\ \bibnamefont {Gudgel}},\
  }\href@noop {} {\bibfield  {journal} {\bibinfo  {journal} {Journal of Crystal
  Growth}\ }\textbf {\bibinfo {volume} {271}},\ \bibinfo {pages} {481}
  (\bibinfo {year} {2004})}\BibitemShut {NoStop}%
\bibitem [{\citenamefont {Jackson}\ \emph {et~al.}(1995)\citenamefont
  {Jackson}, \citenamefont {Gilmer},\ and\ \citenamefont
  {Temkin}}]{jackson1995monte}%
  \BibitemOpen
  \bibfield  {author} {\bibinfo {author} {\bibfnamefont {K.~A.}\ \bibnamefont
  {Jackson}}, \bibinfo {author} {\bibfnamefont {G.~H.}\ \bibnamefont {Gilmer}},
  \ and\ \bibinfo {author} {\bibfnamefont {D.~E.}\ \bibnamefont {Temkin}},\
  }\href@noop {} {\bibfield  {journal} {\bibinfo  {journal} {Physical Review
  Letters}\ }\textbf {\bibinfo {volume} {75}},\ \bibinfo {pages} {2530}
  (\bibinfo {year} {1995})}\BibitemShut {NoStop}%
\bibitem [{\citenamefont {Pawley}\ \emph {et~al.}(1984)\citenamefont {Pawley},
  \citenamefont {Swendsen}, \citenamefont {Wallace},\ and\ \citenamefont
  {Wilson}}]{pawley1984monte}%
  \BibitemOpen
  \bibfield  {author} {\bibinfo {author} {\bibfnamefont {G.}~\bibnamefont
  {Pawley}}, \bibinfo {author} {\bibfnamefont {R.}~\bibnamefont {Swendsen}},
  \bibinfo {author} {\bibfnamefont {D.}~\bibnamefont {Wallace}}, \ and\
  \bibinfo {author} {\bibfnamefont {K.}~\bibnamefont {Wilson}},\ }\href@noop {}
  {\bibfield  {journal} {\bibinfo  {journal} {Physical Review B}\ }\textbf
  {\bibinfo {volume} {29}},\ \bibinfo {pages} {4030} (\bibinfo {year}
  {1984})}\BibitemShut {NoStop}%
\bibitem [{\citenamefont {Whitelam}\ \emph {et~al.}(2014)\citenamefont
  {Whitelam}, \citenamefont {Hedges},\ and\ \citenamefont
  {Schmit}}]{whitelam2014self}%
  \BibitemOpen
  \bibfield  {author} {\bibinfo {author} {\bibfnamefont {S.}~\bibnamefont
  {Whitelam}}, \bibinfo {author} {\bibfnamefont {L.~O.}\ \bibnamefont
  {Hedges}}, \ and\ \bibinfo {author} {\bibfnamefont {J.~D.}\ \bibnamefont
  {Schmit}},\ }\href@noop {} {\bibfield  {journal} {\bibinfo  {journal}
  {Physical Review Letters}\ }\textbf {\bibinfo {volume} {112}},\ \bibinfo
  {pages} {155504} (\bibinfo {year} {2014})}\BibitemShut {NoStop}%
\bibitem [{\citenamefont {Ryu}\ and\ \citenamefont
  {Cai}(2010{\natexlab{a}})}]{ryu2010numerical}%
  \BibitemOpen
  \bibfield  {author} {\bibinfo {author} {\bibfnamefont {S.}~\bibnamefont
  {Ryu}}\ and\ \bibinfo {author} {\bibfnamefont {W.}~\bibnamefont {Cai}},\
  }\href@noop {} {\bibfield  {journal} {\bibinfo  {journal} {Physical Review
  E}\ }\textbf {\bibinfo {volume} {82}},\ \bibinfo {pages} {011603} (\bibinfo
  {year} {2010}{\natexlab{a}})}\BibitemShut {NoStop}%
\bibitem [{\citenamefont {Ryu}\ and\ \citenamefont
  {Cai}(2010{\natexlab{b}})}]{ryu2010validity}%
  \BibitemOpen
  \bibfield  {author} {\bibinfo {author} {\bibfnamefont {S.}~\bibnamefont
  {Ryu}}\ and\ \bibinfo {author} {\bibfnamefont {W.}~\bibnamefont {Cai}},\
  }\href@noop {} {\bibfield  {journal} {\bibinfo  {journal} {Physical Review
  E}\ }\textbf {\bibinfo {volume} {81}},\ \bibinfo {pages} {030601} (\bibinfo
  {year} {2010}{\natexlab{b}})}\BibitemShut {NoStop}%
\bibitem [{\citenamefont {Shneidman}\ \emph {et~al.}(1999)\citenamefont
  {Shneidman}, \citenamefont {Jackson},\ and\ \citenamefont
  {Beatty}}]{Shneidman:1999}%
  \BibitemOpen
  \bibfield  {author} {\bibinfo {author} {\bibfnamefont {V.~A.}\ \bibnamefont
  {Shneidman}}, \bibinfo {author} {\bibfnamefont {K.~A.}\ \bibnamefont
  {Jackson}}, \ and\ \bibinfo {author} {\bibfnamefont {K.~M.}\ \bibnamefont
  {Beatty}},\ }\href@noop {} {\bibfield  {journal} {\bibinfo  {journal} {The
  Journal of Chemical Physics}\ }\textbf {\bibinfo {volume} {111}},\ \bibinfo
  {pages} {6932} (\bibinfo {year} {1999})}\BibitemShut {NoStop}%
\bibitem [{Note1()}]{Note1}%
  \BibitemOpen
  \bibinfo {note} {Here $\sigma \equiv (\sigma _\parallel + \sigma _{\protect
  \rm diag})/(2\protect \sqrt {\chi })$, with $\sigma _\parallel \equiv 2J -
  \protect \qopname \relax o{ln}\protect \qopname \relax o{coth}J$, $\sigma
  _{\protect \rm diag} \equiv \protect \sqrt {2} \protect \qopname \relax
  o{ln}\protect \qopname \relax o{sinh}2 J$, and $\chi \equiv (1-\protect
  \qopname \relax o{sinh}^{-4} 2J)^{1/8}$}\BibitemShut {NoStop}%
\bibitem [{\citenamefont {Hedges}\ and\ \citenamefont
  {Whitelam}(2012)}]{hedges2012patterning}%
  \BibitemOpen
  \bibfield  {author} {\bibinfo {author} {\bibfnamefont {L.~O.}\ \bibnamefont
  {Hedges}}\ and\ \bibinfo {author} {\bibfnamefont {S.}~\bibnamefont
  {Whitelam}},\ }\href@noop {} {\bibfield  {journal} {\bibinfo  {journal} {Soft
  Matter}\ }\textbf {\bibinfo {volume} {8}},\ \bibinfo {pages} {8624} (\bibinfo
  {year} {2012})}\BibitemShut {NoStop}%
\bibitem [{Note2()}]{Note2}%
  \BibitemOpen
  \bibinfo {note} {If the cross-sectional area $N_\perp \equiv N_x \times N_y$
  of the simulation box is too small to accommodate the 2D critical cluster,
  $N_\perp < N_{\protect \rm c}$, then (\ref {barrier}) should be replaced by
  $G(N_\perp )$.}\BibitemShut {Stop}%
\bibitem [{\citenamefont {Jack}\ \emph {et~al.}(2007)\citenamefont {Jack},
  \citenamefont {Hagan},\ and\ \citenamefont {Chandler}}]{jack2007fluctuation}%
  \BibitemOpen
  \bibfield  {author} {\bibinfo {author} {\bibfnamefont {R.~L.}\ \bibnamefont
  {Jack}}, \bibinfo {author} {\bibfnamefont {M.~F.}\ \bibnamefont {Hagan}}, \
  and\ \bibinfo {author} {\bibfnamefont {D.}~\bibnamefont {Chandler}},\
  }\href@noop {} {\bibfield  {journal} {\bibinfo  {journal} {Physical Review
  E}\ }\textbf {\bibinfo {volume} {76}},\ \bibinfo {pages} {021119} (\bibinfo
  {year} {2007})}\BibitemShut {NoStop}%
\bibitem [{\citenamefont {Leamy}\ \emph {et~al.}(1980)\citenamefont {Leamy},
  \citenamefont {Bean}, \citenamefont {Poate},\ and\ \citenamefont
  {Celler}}]{leamy1980nonequilibrium}%
  \BibitemOpen
  \bibfield  {author} {\bibinfo {author} {\bibfnamefont {H.~J.}\ \bibnamefont
  {Leamy}}, \bibinfo {author} {\bibfnamefont {J.~C.}\ \bibnamefont {Bean}},
  \bibinfo {author} {\bibfnamefont {J.}~\bibnamefont {Poate}}, \ and\ \bibinfo
  {author} {\bibfnamefont {G.}~\bibnamefont {Celler}},\ }\href@noop {}
  {\bibfield  {journal} {\bibinfo  {journal} {Journal of Crystal Growth}\
  }\textbf {\bibinfo {volume} {48}},\ \bibinfo {pages} {379} (\bibinfo {year}
  {1980})}\BibitemShut {NoStop}%
\bibitem [{\citenamefont {Kim}\ \emph {et~al.}(2008)\citenamefont {Kim},
  \citenamefont {Scarlett}, \citenamefont {Biancaniello}, \citenamefont
  {Sinno},\ and\ \citenamefont {Crocker}}]{kim2008probing}%
  \BibitemOpen
  \bibfield  {author} {\bibinfo {author} {\bibfnamefont {A.}~\bibnamefont
  {Kim}}, \bibinfo {author} {\bibfnamefont {R.}~\bibnamefont {Scarlett}},
  \bibinfo {author} {\bibfnamefont {P.}~\bibnamefont {Biancaniello}}, \bibinfo
  {author} {\bibfnamefont {T.}~\bibnamefont {Sinno}}, \ and\ \bibinfo {author}
  {\bibfnamefont {J.}~\bibnamefont {Crocker}},\ }\href@noop {} {\bibfield
  {journal} {\bibinfo  {journal} {Nature materials}\ }\textbf {\bibinfo
  {volume} {8}},\ \bibinfo {pages} {52} (\bibinfo {year} {2008})}\BibitemShut
  {NoStop}%
\bibitem [{\citenamefont {Van~Siclen}\ and\ \citenamefont
  {Wolfer}(1992)}]{van1992nonequilibrium}%
  \BibitemOpen
  \bibfield  {author} {\bibinfo {author} {\bibfnamefont {C.~D.}\ \bibnamefont
  {Van~Siclen}}\ and\ \bibinfo {author} {\bibfnamefont {W.}~\bibnamefont
  {Wolfer}},\ }\href@noop {} {\bibfield  {journal} {\bibinfo  {journal} {Acta
  metallurgica et materialia}\ }\textbf {\bibinfo {volume} {40}},\ \bibinfo
  {pages} {2091} (\bibinfo {year} {1992})}\BibitemShut {NoStop}%
\bibitem [{\citenamefont {Berthier}\ \emph {et~al.}(2017)\citenamefont
  {Berthier}, \citenamefont {Charbonneau}, \citenamefont {Flenner},\ and\
  \citenamefont {Zamponi}}]{berthier2017create}%
  \BibitemOpen
  \bibfield  {author} {\bibinfo {author} {\bibfnamefont {L.}~\bibnamefont
  {Berthier}}, \bibinfo {author} {\bibfnamefont {P.}~\bibnamefont
  {Charbonneau}}, \bibinfo {author} {\bibfnamefont {E.}~\bibnamefont
  {Flenner}}, \ and\ \bibinfo {author} {\bibfnamefont {F.}~\bibnamefont
  {Zamponi}},\ }\href@noop {} {\bibfield  {journal} {\bibinfo  {journal} {arXiv
  preprint arXiv:1706.02738}\ } (\bibinfo {year} {2017})}\BibitemShut {NoStop}%
\bibitem [{\citenamefont {Corezzi}\ \emph {et~al.}(2009)\citenamefont
  {Corezzi}, \citenamefont {De~Michele}, \citenamefont {Zaccarelli},
  \citenamefont {Tartaglia},\ and\ \citenamefont
  {Sciortino}}]{corezzi2009connecting}%
  \BibitemOpen
  \bibfield  {author} {\bibinfo {author} {\bibfnamefont {S.}~\bibnamefont
  {Corezzi}}, \bibinfo {author} {\bibfnamefont {C.}~\bibnamefont {De~Michele}},
  \bibinfo {author} {\bibfnamefont {E.}~\bibnamefont {Zaccarelli}}, \bibinfo
  {author} {\bibfnamefont {P.}~\bibnamefont {Tartaglia}}, \ and\ \bibinfo
  {author} {\bibfnamefont {F.}~\bibnamefont {Sciortino}},\ }\href@noop {}
  {\bibfield  {journal} {\bibinfo  {journal} {The Journal of Physical Chemistry
  B}\ }\textbf {\bibinfo {volume} {113}},\ \bibinfo {pages} {1233} (\bibinfo
  {year} {2009})}\BibitemShut {NoStop}%
\bibitem [{\citenamefont {Whitelam}\ \emph {et~al.}(2012)\citenamefont
  {Whitelam}, \citenamefont {Schulman},\ and\ \citenamefont
  {Hedges}}]{whitelam2012self}%
  \BibitemOpen
  \bibfield  {author} {\bibinfo {author} {\bibfnamefont {S.}~\bibnamefont
  {Whitelam}}, \bibinfo {author} {\bibfnamefont {R.}~\bibnamefont {Schulman}},
  \ and\ \bibinfo {author} {\bibfnamefont {L.}~\bibnamefont {Hedges}},\
  }\href@noop {} {\bibfield  {journal} {\bibinfo  {journal} {Physical Review
  Letters}\ }\textbf {\bibinfo {volume} {109}},\ \bibinfo {pages} {265506}
  (\bibinfo {year} {2012})}\BibitemShut {NoStop}%
\bibitem [{\citenamefont {Ke}\ \emph {et~al.}(2012)\citenamefont {Ke},
  \citenamefont {Ong}, \citenamefont {Shih},\ and\ \citenamefont
  {Yin}}]{ke2012three}%
  \BibitemOpen
  \bibfield  {author} {\bibinfo {author} {\bibfnamefont {Y.}~\bibnamefont
  {Ke}}, \bibinfo {author} {\bibfnamefont {L.~L.}\ \bibnamefont {Ong}},
  \bibinfo {author} {\bibfnamefont {W.~M.}\ \bibnamefont {Shih}}, \ and\
  \bibinfo {author} {\bibfnamefont {P.}~\bibnamefont {Yin}},\ }\href@noop {}
  {\bibfield  {journal} {\bibinfo  {journal} {Science}\ }\textbf {\bibinfo
  {volume} {338}},\ \bibinfo {pages} {1177} (\bibinfo {year}
  {2012})}\BibitemShut {NoStop}%
\bibitem [{\citenamefont {Hasenbusch}\ \emph {et~al.}(1996)\citenamefont
  {Hasenbusch}, \citenamefont {Meyer},\ and\ \citenamefont
  {P{\"u}tz}}]{hasenbusch1996roughening}%
  \BibitemOpen
  \bibfield  {author} {\bibinfo {author} {\bibfnamefont {M.}~\bibnamefont
  {Hasenbusch}}, \bibinfo {author} {\bibfnamefont {S.}~\bibnamefont {Meyer}}, \
  and\ \bibinfo {author} {\bibfnamefont {M.}~\bibnamefont {P{\"u}tz}},\
  }\href@noop {} {\bibfield  {journal} {\bibinfo  {journal} {Journal of
  statistical physics}\ }\textbf {\bibinfo {volume} {85}},\ \bibinfo {pages}
  {383} (\bibinfo {year} {1996})}\BibitemShut {NoStop}%
\bibitem [{\citenamefont {Fredrickson}\ and\ \citenamefont
  {Andersen}(1984)}]{fredrickson1984kinetic}%
  \BibitemOpen
  \bibfield  {author} {\bibinfo {author} {\bibfnamefont {G.~H.}\ \bibnamefont
  {Fredrickson}}\ and\ \bibinfo {author} {\bibfnamefont {H.~C.}\ \bibnamefont
  {Andersen}},\ }\href@noop {} {\bibfield  {journal} {\bibinfo  {journal}
  {Physical Review Letters}\ }\textbf {\bibinfo {volume} {53}},\ \bibinfo
  {pages} {1244} (\bibinfo {year} {1984})}\BibitemShut {NoStop}%
\end{thebibliography}

%

\clearpage

\onecolumngrid

\renewcommand{\theequation}{A\arabic{equation}}
\renewcommand{\thefigure}{A\arabic{figure}}
\renewcommand{\thesection}{A\arabic{section}}

\setcounter{equation}{0}
\setcounter{section}{0}
\setcounter{figure}{0}

\setlength{\parskip}{0.25cm}%
\setlength{\parindent}{0pt}%

\appendix

\section{Approximation of the growth front as a 2D Ising model}
\label{app_a}
For $\epsilon \gtrsim 1.630$ the equilibrium interface between particles and vacancies is statistically smooth\c{hasenbusch1996roughening}. For sufficiently large $\epsilon$ $(\gtrsim 2)$ it is convenient to consider the exposed surface of a particle structure growing in the $z$-direction to be a two-dimensional (2D) Ising model\c{burton1951growth}. If the layer adjacent to the exposed surface has no vacancies then the exposed layer behaves as a 2D Ising model whose parameters are the same as the 3D Ising model given in the main text, $J=\epsilon/4$ and $h=\Delta g/2$. To see this, note that the Hamiltonian of the exposed layer is 
\beq
{\cal H}= -\epsilon \sum_{<ij>} n_i n_j + \mu \sum_i n_i - \epsilon \sum_i n_i,
\eeq
where $n_i=1 (0)$ for a particle (vacancy). The first sum runs over all distinct pairs of in-plane bonds, and the second and third sums run over all in-plane sites. The last term accounts for bonds between the exposed layer and the layer below (which we assume to be perfect, with no vacancies). Setting $n_i = (S_i + 1)/2$ gives, up to constant terms,
\beq
\label{hamil}
{\cal H}= -\frac{\epsilon}{4} \sum_{<ij>} S_i S_j + \frac{1}{2} \left[ \mu - \epsilon \left(1+\frac{z_{\rm p}}{2}\right)\right] \sum_i S_i,
\eeq
where $z_{\rm p}=4$ is the in-plane coordination number. \eqq{hamil} is the Ising Hamiltonian with $J=\epsilon/4$ and $h=-\left( \mu - 3 \epsilon\right)/2=\Delta g/2$.

The 2D Ising critical temperate corresponds to a value $\epsilon \approx 1.762$\c{onsager1944crystal}. Thus if we approximate the surface of the structure as a 2D Ising model, then, for $\epsilon \gtrsim 1.762$, there exists a stable interface (a positive line tension) between particles and vacancies in 2D. Successive layers of the three-dimensional structure face a free-energy barrier to their formation, and a 3D structure can grow in a layer-by-layer manner, with successive 2D nucleation events required for advance of the growth front\c{gilmer1976growth}.

\section{Internal relaxation of the bulk}
\label{app_vacancies}

Vacancies trapped within the structure can undergo diffusion, in an effective way, even in the presence of the kinetic constraint: the particle adjacent to the vacancy, which has fewer than 6 neighbors, can become a vacancy, and then the original vacancy can be filled in. In addition, two vacancies that meet each other can coalesce, leaving behind only a single vacancy. This internal dynamics of (effective) vacancy diffusion and coalescence is similar to that of spins in the kinetically constrained Fredrickson-Andersen model\c{fredrickson1984kinetic}. Vacancy coalescence can lead to evolution of the bulk structure toward the equilibrium vacancy density, which we see for sufficiently small values of $\epsilon$ ($\lesssim 2$): see \f{fig_relax}(c). By contrast, for $\epsilon \gtrsim 2$, the vacancy density $\phi$ is independent of observation time, for the range of times studied, showing that no aging of the structure has occurred on the growth timescale. Thus the dynamically-generated vacancy density results only from dynamics that occurs in the presence of the growth front, and not from subsequent relaxation of the bulk of the structure. Vacancy coalescence is unphysical in the sense that it would not happen within the bulk of a solid structure, and so we focus our attention on the regime of parameter space in which this process does not occur.

\clearpage
\section{Additional figures}

\begin{figure*}[ht] 
   \centering
  \includegraphics[width=\linewidth]{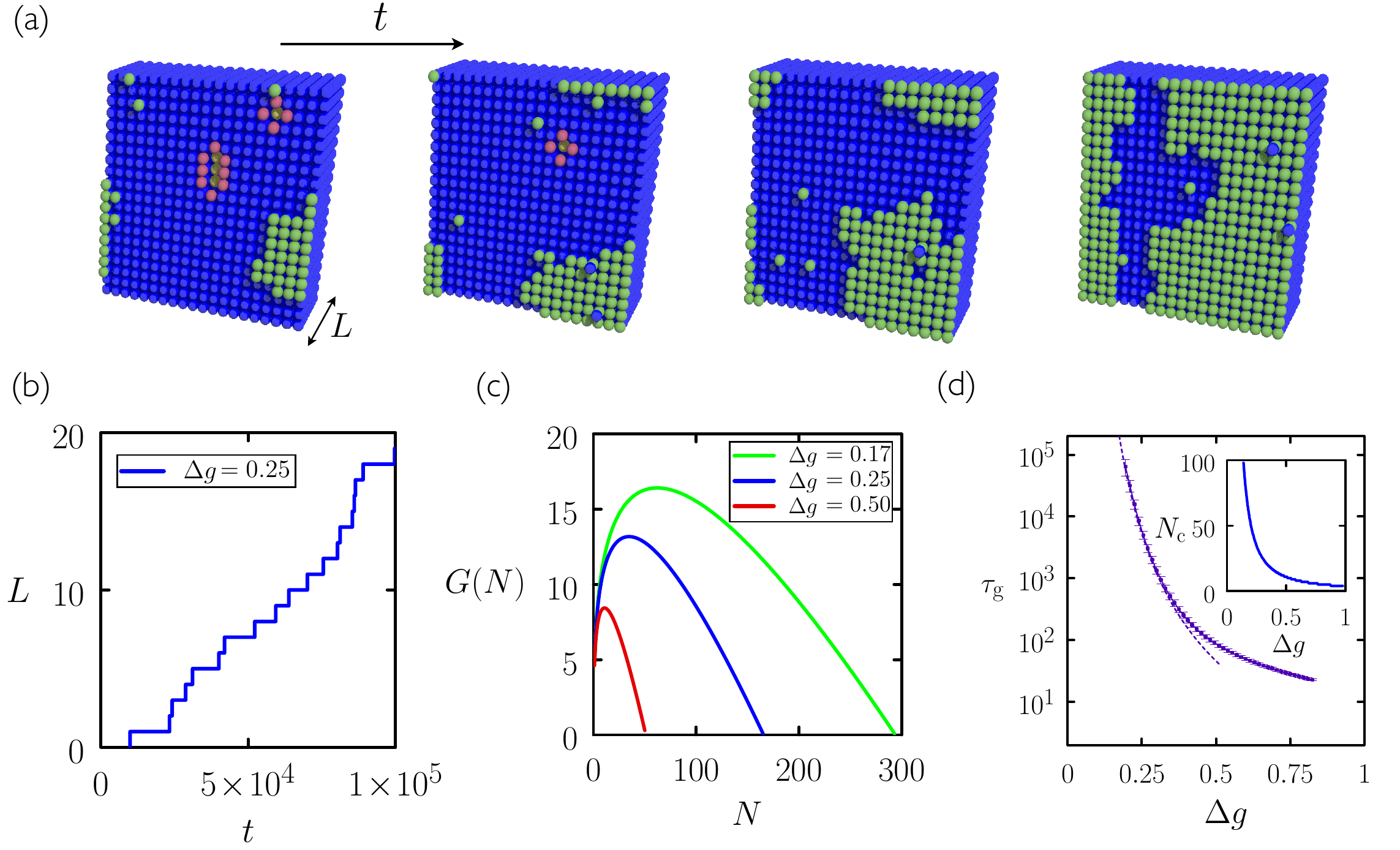} 
   \caption{Layer-by-layer growth in the 3D lattice gas. (a) Time-ordered configurations of a simulation box of $20 \times 20 \times 50$ lattice sites, for parameters $\epsilon=2.55, \Delta g =0.25$ (see e.g. Refs.\c{gilmer1976growth,gilmer1980computer} for similar pictures). $L$ is the number of layers grown in the $z$-direction; periodic boundaries are applied in the direction perpendicular to growth. Particles are shown blue, with the following exceptions: particles in the nucleating layer are shown green; under-coordinated particles in the layer below that are shown pink; and the exposed particles in the layer below that are shown yellow. The first snapshot shows a critical 2D cluster (green). (b) A plot of $L$ versus $t$ shows that the interface pauses, for varying amounts of time (number of Monte Carlo sweeps), between 2D nucleation events. (c) These nucleation events are governed by free-energy profiles, \eqq{ryu}, for 2D clusters on the surface of the 3D structure. We show such profiles, as a function of cluster size $N$, for three values of $\Delta g$. These profiles assume that the layer adjacent to the nucleating layer is defect-free, which is approximately true for large $\epsilon$. (d) Characteristic growth time $\tau_{\rm g}$ as a function of $\Delta g$ (blue line). Overlaid as a dashed line is \eqq{eq_tau}, showing that, for sufficiently small $\Delta g$, the scaling of growth time follows from consideration of 2D nucleation events. The inset shows the size $N_c$ of the 2D critical cluster as a function of $\Delta g$.}
   \label{fig_s1}
   \end{figure*}
   
   \clearpage
   
   \begin{figure}
   \centering
  \includegraphics[width=0.6\linewidth]{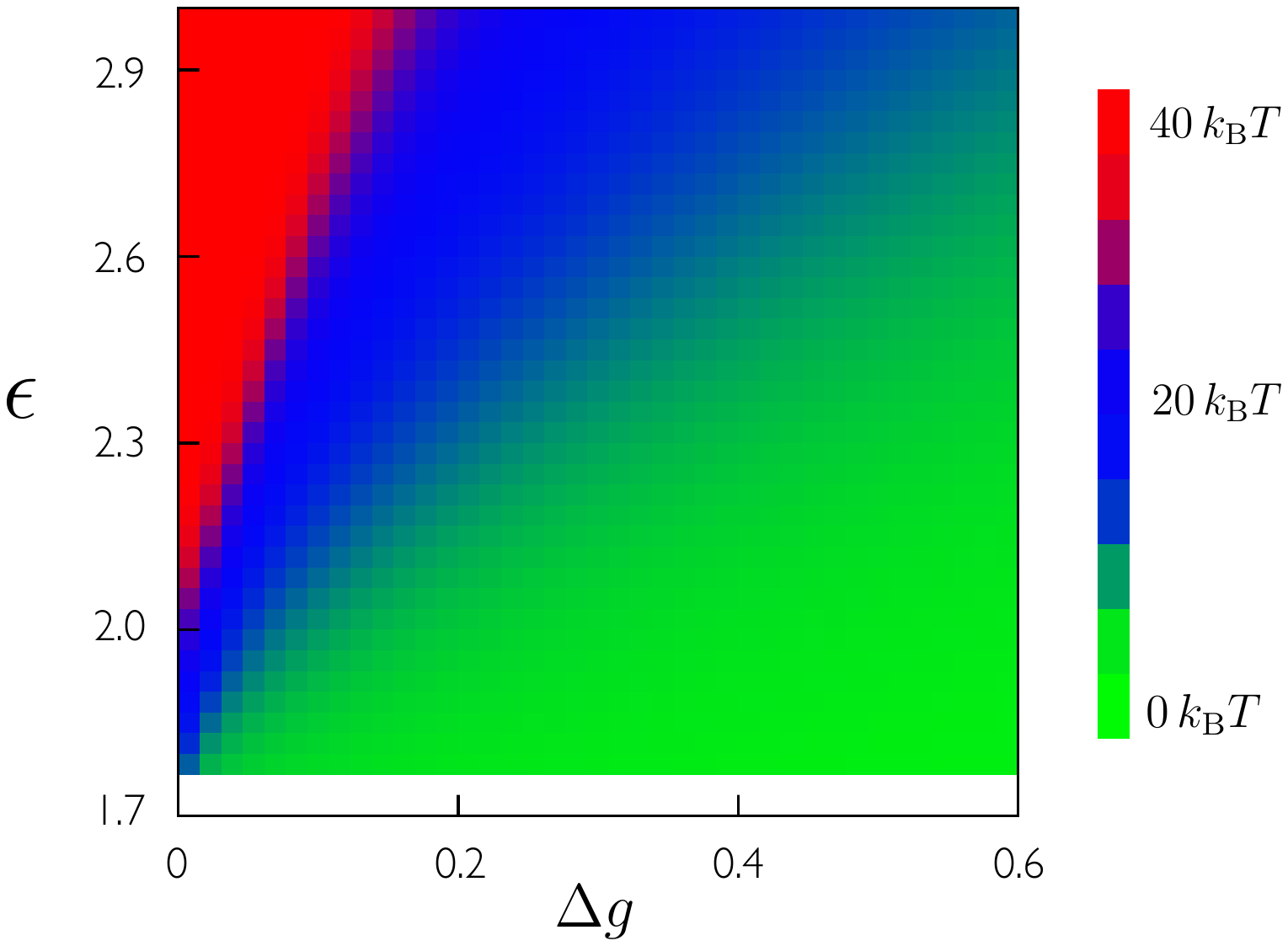} 
   \caption{Free-energy barrier, \eqq{barrier}, to nucleation of a 2D layer on the surface of a defect-free 3D structure. The barrier increases with decreasing $\Delta g$ or increasing $\epsilon$. Its increase with $\epsilon$ is approximately quadratic. Thus the growth time of the 3D structure, roughly the exponential of the barrier, grows faster with $\epsilon$ than the molecular relaxation time, which scales exponentially with $\epsilon$.}
   \label{fig_barrier}
\end{figure}

\clearpage

\begin{figure*}[] 
   \centering
  \includegraphics[width=\linewidth]{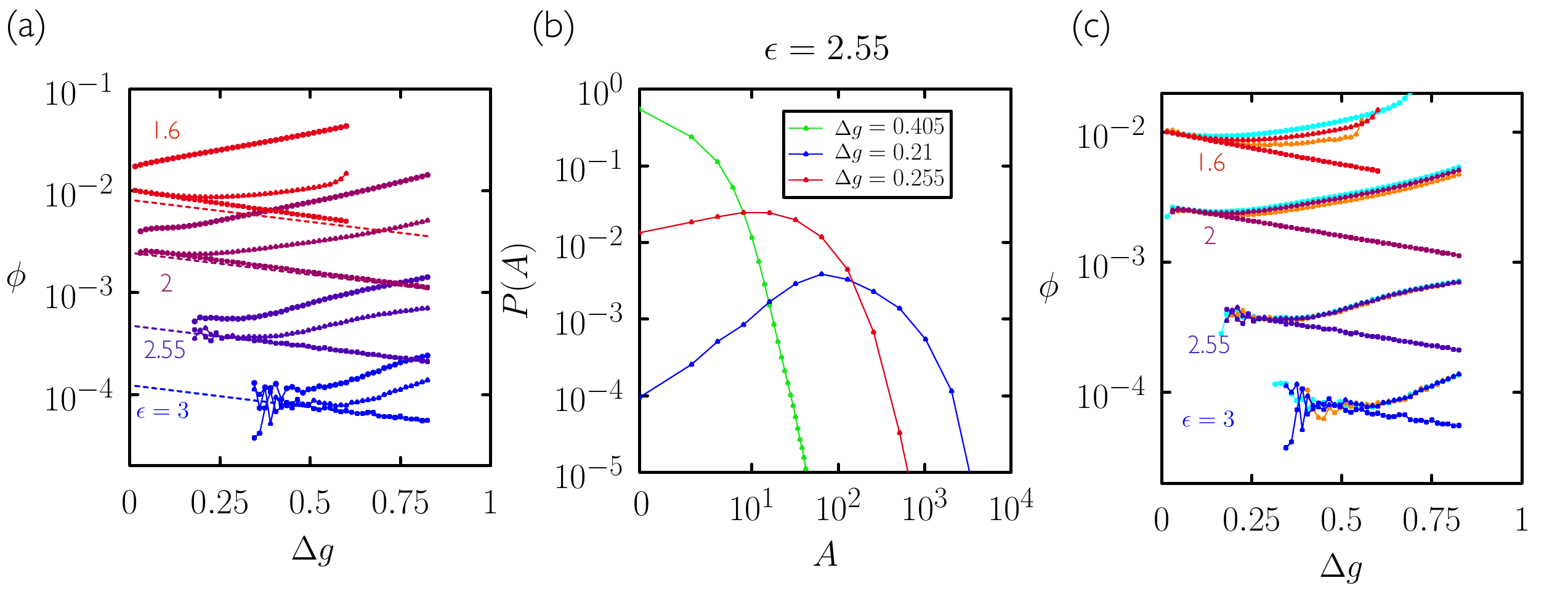} 
   \caption{(a) As \f{fig1}(b), but including data (upwards-sloping lines with circles) indicating the vacancy density of the `fresh bulk', i.e. the fraction of sites that are vacant upon first acquiring 6 neighbors. (b) Probability distribution $P(A)$ of the number of times $A$ that a site changes state after first acquiring 6 neighbors, for three values of $\Delta g$ and for $\epsilon=2.55$. Such changes of state allow the `fresh bulk' adjacent to the growth front to evolve into the `mature bulk' (triangle symbols in panel (a)). For small values of $\Delta g$ such evolution is sufficient to attain equilibrium while a site is close to the growth front; for large values of $\Delta g$ it is not. (Here and in \f{fig_scale} we show even values of $A$; histograms for odd values of $A$ show similar behavior.) (c) As \f{fig1}(b) but with additional dynamic data: light blue and orange lines show the vacancy density $\phi$ immediately after the growth of 25 and 100 layers, respectively (the data of \f{fig1}(b) are obtained immediately after the growth of 50 layers). For the cases $\epsilon=1.6$ and 2, the dynamically-generated vacancy density depends on the observation time, because vacancy-vacancy coalescence within the growing structure causes some relaxation within the bulk of the structure as it grows. For the other two cases the blue and orange lines are not visible (they lie under the original sets of data), showing that bulk relaxation does not operate on the timescale of growth (the vacancy density is too low and encounters between vacancies too rare). Consequently, the nonequilibrium vacancy density results from the dynamical processes associated with creation and advance of the growth front, leading to the scaling behavior seen in \f{fig_scale}(c).}
   \label{fig_relax}
\end{figure*}

\clearpage

\begin{figure*}[t]
   \centering
  \includegraphics[width=0.85\linewidth]{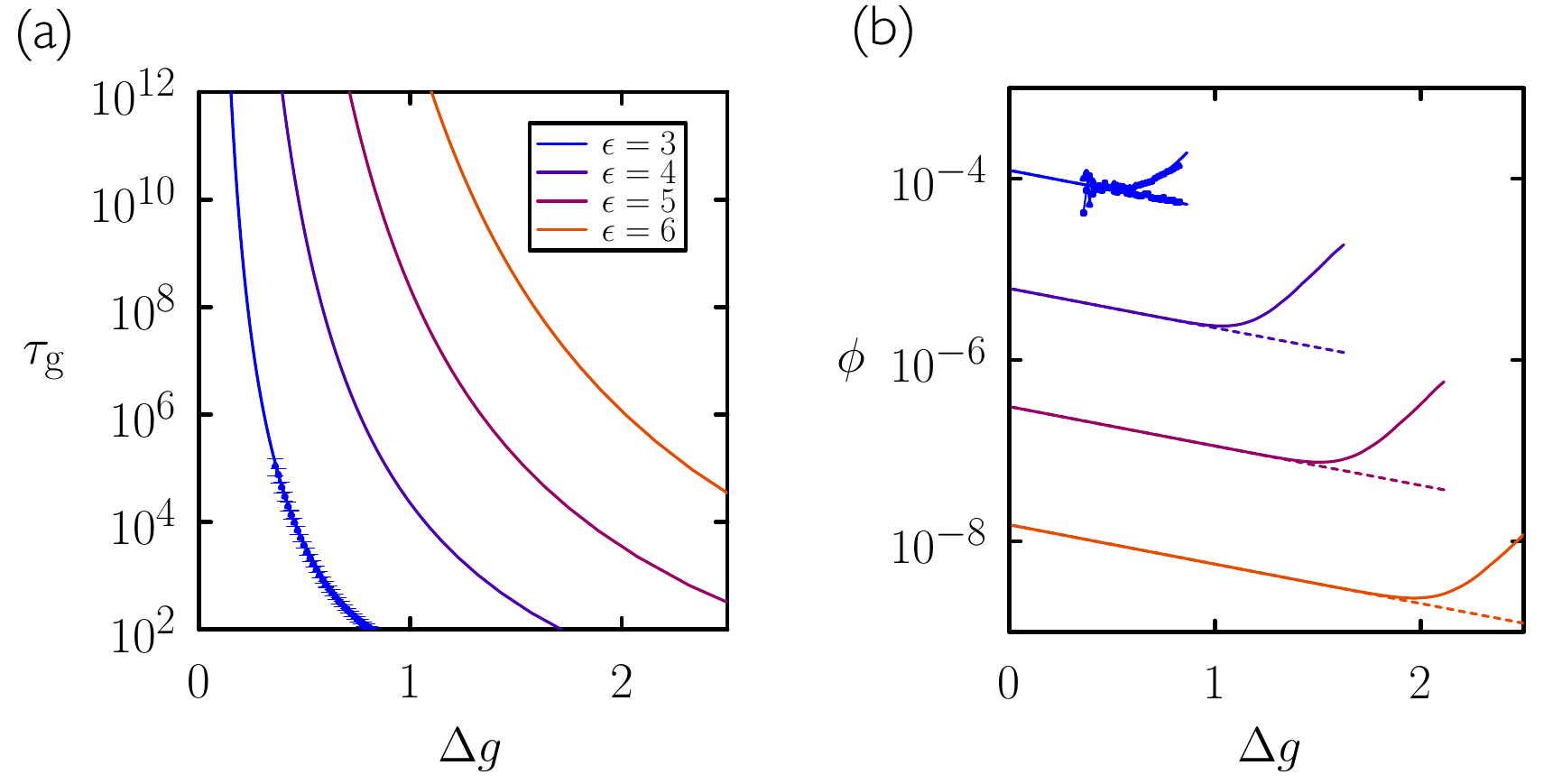} 
   \caption{As \f{fig1}, but extrapolated to larger values of $\epsilon$ (longer times and smaller impurity densities) using \eqq{scaling}. For $\epsilon=3$ we also show simulation data.}
   \label{fig1_extrap}
\end{figure*}

\end{document}